\documentclass[runningheads]{llncs}

\usepackage{amsmath,amssymb,amsfonts}
\usepackage[T1]{fontenc}
\usepackage{algorithmic}
\usepackage{graphicx}
\usepackage{textcomp}
\usepackage[square,numbers]{natbib}
\usepackage[caption=false]{subfig}
\usepackage{xcolor}

 \usepackage{balance} 
\def\BibTeX{{\rm B\kern-.05em{\sc i\kern-.025em b}\kern-.08em
    T\kern-.1667em\lower.7ex\hbox{E}\kern-.125emX}}
    
\begin{document}
\title{Detection of ransomware attacks using federated learning based on the CNN model}
%
%
\author{Hong Nhung Nguyen${}^{1}$, Ha Thanh Nguyen${}^{2}$, Damien Lescos${}^{3}$}
\authorrunning{Nguyen et al.}
%
\institute{Gachon University, Korea \and
National Institute of Informatics, Japan \and
SitinCloud, France}
\maketitle              
\begin{abstract}
Computing is still under a significant threat from ransomware, which necessitates prompt action to prevent it. Ransomware attacks can have a negative impact on how smart grids, particularly digital substations. In addition to examining a ransomware detection method using artificial intelligence (AI), this paper offers a ransomware attack modeling technique that targets the disrupted operation of a digital substation. The first, binary data is transformed into image data and fed into the convolution neural network model using federated learning. The experimental findings demonstrate that the suggested technique detects ransomware with a high accuracy rate.

\end{abstract}
\section{Introduction}
Nowadays, cyber threats are one of the costliest losses an institution can encounter. Ransomware is malware that threatens to publish the victim's data or block access to it unless a ransom is paid. It became popular in the early 2010s, and its use has been overgrown. Ransomware attacks have been increasing in number and sophistication in recent years. WannaCry, for example, was a ransomware worm that spread rapidly across the world in May 2017, affecting more than 230,000 computers in 150 countries.

With the development of federated learning techniques, it becomes possible to train machine learning models on data that is distributed across different devices or organizations. The federated learning approach has the potential to be used in training models for detecting ransomware. The reason is that, in a federated learning setting, each data owner keeps its data locally and only shares model updates with a central server. Therefore, the data never leaves the owner's premises, which alleviates privacy concerns.

A convolutional neural network (CNN) is a type of deep learning neural network that is generally used for image classification and recognition. It has been shown to be effective in various types of image classification tasks. An idea comes to using the CNN model to classify ransomware attacks. The advantages of using the CNN model are that it can automatically learn features from data and that it is robust to data variability. In addition, it also allows the representation of the data in a more compact and efficient way.

The paper first preprocesses the data into images and then uses the CNN model to learn and classify the data. The proposed method has the advantage of being privacy-preserving because the data never leaves the data owner's premises. By experimenting, we found that our proposed model is more accurate than several traditional methods in the literature.

There are several methods proposed in the literature for detecting ransomware. Takeuchi et al. proposed a detection method using support vector machines (SVMs). The key idea is to trace the API call when the ransomware is executing. With this approach, the authors can detect unseen ransomware by the similarity of API calls between samples. Another approach proposed by Arabo et al. is to monitor the behavior of the ransomware process. The key idea is to monitor the process behavior to detect ransomware. In this approach, the authors look into the key process usages to detect ransomware. They also include the analysis of DLLs and system calls.

However, the ransomware detection methods proposed in the literature have several limitations. First, many of them require the use of static features, which may be hard to collect, especially on obfuscated and metamorphic binaries. Second, many of them require the use of dynamic features, which means that they require the execution of the ransomware for analysis. As a result, we may want some approximate static features that can be used to detect ransomware.

\section{Background}

In this section, some key concepts related to Machine learning and cyber security are discussed in order to understand and appreciate the novelties of the proposed approach.

\subsection{Machine Learning for Cyber Security}

\begin{figure}[t]
    \centering
    \includegraphics[width=.7\textwidth]{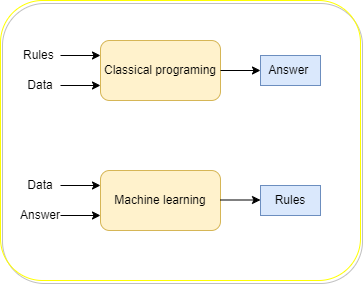}
    \caption{Machine learning: a new programming paradigm}
    \label{machinelearning}
\end{figure}

Currently, we can apply machine learning to solve issues in the real world, such as computer vision and natural language processing. Using the technique of machine learning (ML), which is a type of artificial intelligence (AI), software programs can make predictions more accurately with having to be explicitly trained to do so \cite{bookpython}.  As shown in Figure \ref{machinelearning}, In the paradigm of symbolic AI known as classical programming, humans input rules (a program) and data to be processed in accordance with these rules, and the results are replies. With machine learning, humans input data along with the predictions made from the data, and the rules are produced as a result. Then, using new data, these rules can be employed to develop new results \cite{bookDL}.

In order to estimate new target values, machine learning algorithms use existing data as input.  The research on machine learning is divided into the following categories

\begin{itemize}
\item Supervise learning
Supervised learning is a staple of many machine learning problems. It involves learning any input for which we already know the target or labels. The applications of near-complete deep learning are usually aimed at: object recognition, voice recognition, image classification, and supervised machine learning. Most of these include two problems, namely classification and regression problems. It can also be extended to several issues such as sequence generation, syntax tree prediction, object detection, and image segmentation. Some famous criteria of supervised learning are face recognition and spam detection
\item Unsupervised learning
Unsupervised learning involves learning data without labels. It is usually aimed at, for example, visualizing data or reducing data noise. Unsupervised learning mainly analyzes the data and is often necessary to understand the data better before solving machine learning problems. In fact, dimensional reduction or clustering are well-known algorithms in unsupervised machine learning.
\item Semi-supervised Learning
Semi-supervised learning is a machine learning technique that involves training using a small amount of labeled data and a large amount of unlabeled data. When unlabeled data is combined with a modest bit of labeled data, learning accuracy can be significantly improved. An expert human or a physical experiment is frequently required for collecting the data labels for the training task. Therefore, the cost of labeling may be prohibitive. As a result, the application of semi-supervised machine learning has a lot of potentials.
\item Reinforcement learning
It was long overlooked, but this branch of learning was recently noticed when Google DeepMind successfully applied it to learning to play the game of Go. In reinforcement learning, an actor receives information about its environment and chooses actions that maximize the reward. For example, a neural network idles into an electromagnetic game screen and outputs in-game actions to maximize its score which can be trained through reinforcement learning. Reinforcement learning is a relatively new field of research and has not had significant practical success outside of games. However, we expect reinforcement learning to be widely applied in many fields. such as self-driving cars, robotics, resource management, and education.
\end{itemize}

Federated Learning, a new machine learning technique, is a form of collaborative learning in which multiple machines train a shared model by sharing local models with each other. The objective of federated learning is to enable the use of collective intelligence to enable all devices to learn a shared model directly from user data and improve the accuracy of individual devices. Federated learning is a type of conventional machine learning that is distributed and decentralized, so it can be used when users do not have to share their data with a central server due to privacy and security concerns.

\begin{figure}[t]
    \centering
    \includegraphics[width=.7\textwidth]{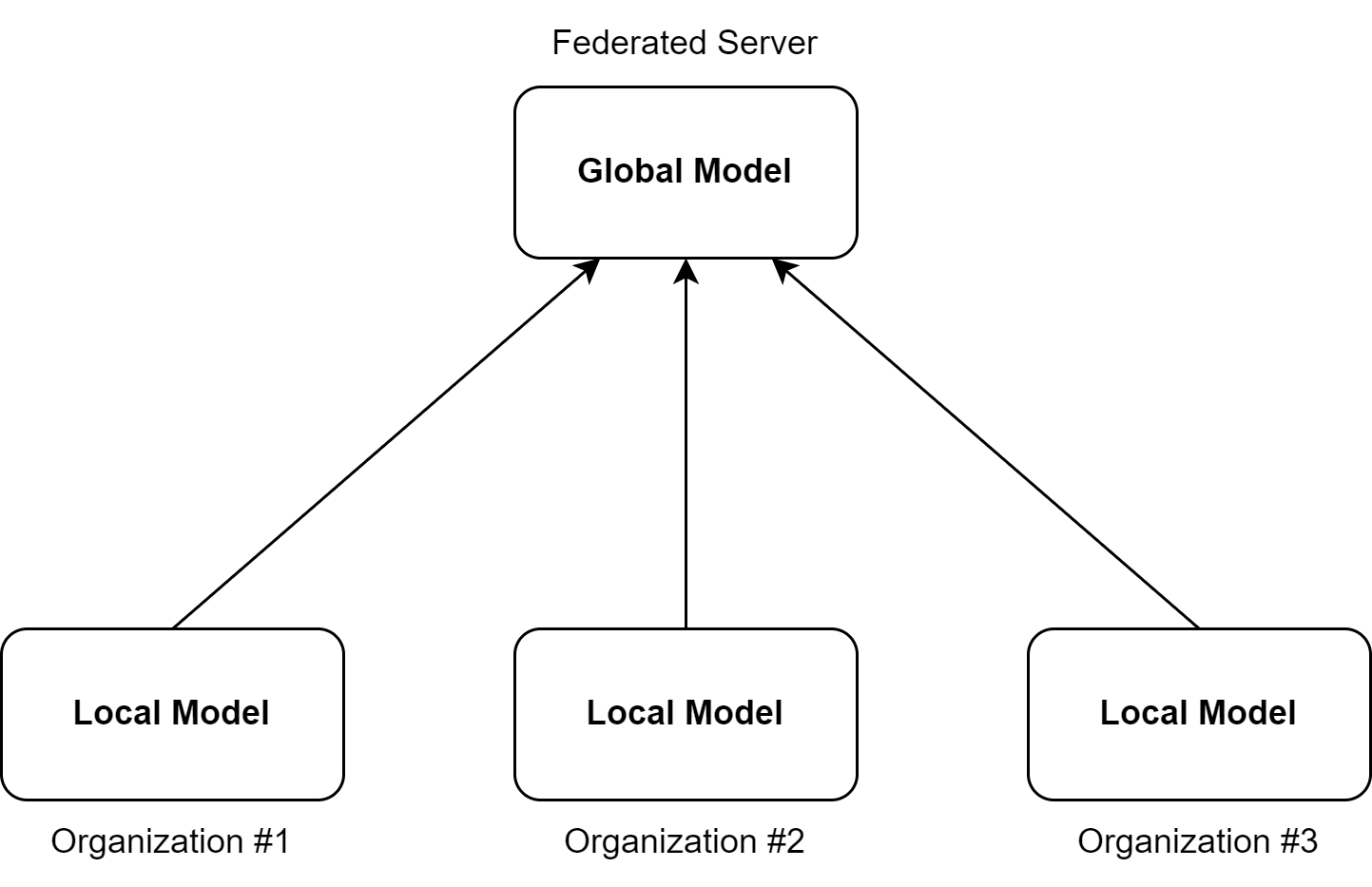}
    \caption{Federated Learning}
    \label{FL}
\end{figure}

As in Figure \ref{FL}, the local models update the global model in federated learning.  The advantage of this method is that the training data is not exposed, and the data privacy is protected. Besides, this technique can be used in many distributed data situations, such as smart health, educational technology, and social networking. 

\subsection{Deep Learning}
Deep learning is a subset of machine learning in which neural networks automatically extract and process data representation. Deep learning can be used to process data in many ways. For example, deep learning can be used to process unordered text data, images, time-series data, and structured data such as relational databases. Deep learning can be used to extract high-level feature representations from data automatically. Deep learning can also be used to learn complex functions from data.

Feed-forward neural network is the most basic type of neural network.  Feed-forward neural networks consist of a series of layers of neurons. The first layer of neurons receives the input data, and each subsequent layer accepts the previous layer's output as input. The last layer produces the output of the neural network.

Recurrent neural networks \cite{john1982neural,hochreiter1997long,cho2014learning} are a type of neural network that is used to model time series data or data that has a sequential nature. In recurrent neural networks, the output of the current layer is fed back to the input of the next layer. This feedback connection allows the recurrent neural network to model the temporal dependence of the input data.

Convolutional neural networks are a type of neural network that originally is used to model data that has a spatial structure. Convolutional neural networks are composed of layers of neurons that are arranged in a three-dimensional grid. In convolutional neural networks, the neurons in each layer are connected to a small region of the previous layer. This connection pattern allows convolutional neural networks to model the spatial structure of the input data.

\section{Proposed Method}
We propose a deep learning model that can classify the given input image according to its class type. We also want to ensure the proposed deep learning model (CNN) can learn the weights and the images properly. We have made presumptions about the techniques we have implemented, and through this section, we want to justify and analyze our approach to this problem more.

\begin{figure*}[t]
    \centering
    \includegraphics[width=.8\textwidth]{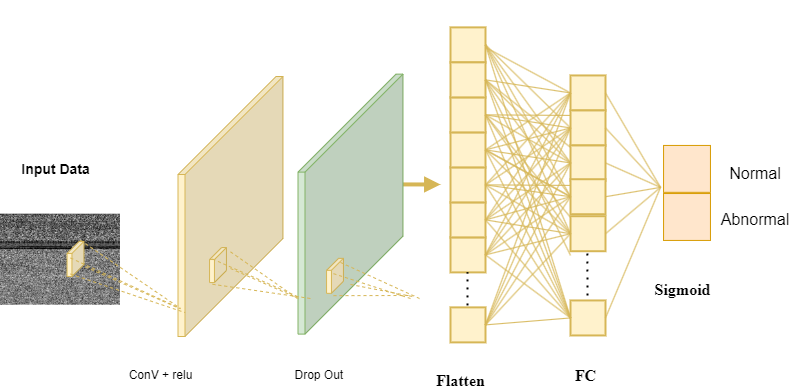}
    \caption{Structure of CNN model}
    \label{cnn}
\end{figure*}
\subsection{Input for the CNN model }

\subsubsection{Data collection}
We initially obtained a vast collection of about 30.000 PE ransomware binaries from \textit{virusshare.com}, collected over the period 2010 to 2017.

Research on the Darknet also provided us with the most recent samples, known for damages caused to vital organisations (for example, the gang behind \textit{Lockbit 3.0} leaked its competitors products).

Ransomware binaries tend to be small and written in low level languages, with notable exceptions (GoLang is trendy among hackers and its runtime creates heavy binaries, even after having been compressed with UPX). That's why about 600 negative examples come from \textit{C:\textbackslash Windows\textbackslash\ System32} as they are small low-level binaries.

The Windows Application Store was used to provide about 2400 others negative examples.

To balance the dataset classes, we keep the 3000 most recent ransomware binaries. Finally, the dataset consists of 3000 positive (malicious) and 3000 negative (benign) examples, in the same order of magnitude as other studies. 

We decide to ignore potential variants among ransomware samples as it would be difficult, if not impossible, to comment on the differences and gains in function that distinguish them.

\subsubsection{Transforming binaries into images}
\begin{figure*}[t]
    \centering
    \includegraphics[width=.8\textwidth]{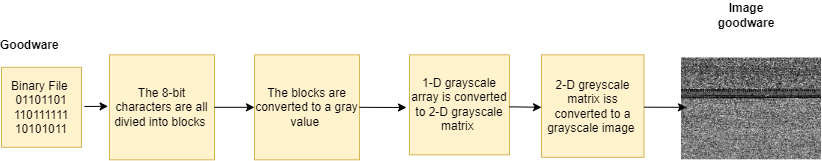}
    \caption{Data Processing}
    \label{convert}
\end{figure*}

Firstly, we collect software from the window files, then convert data into image format, as shown in Figure a. The grey image has a size of 300x300 in height and width. The process of converting data into image format is shown in Figure \ref{convert}.

\begin{figure}[ht]
	\centering
	\subfloat[a]{%
		\includegraphics[width=.4\linewidth]{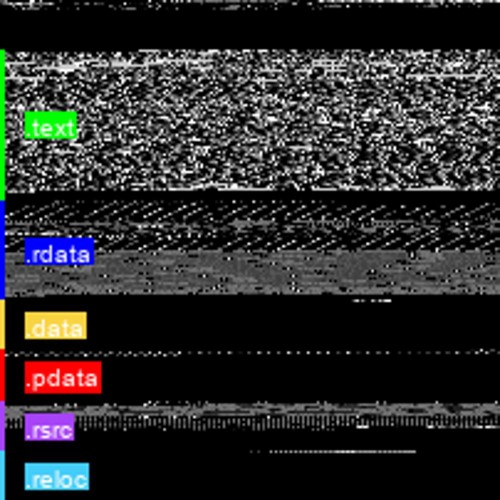}
	\label{Benign binary}} 
	\centering
	\subfloat[b]{%
		\includegraphics[width=.4\linewidth]{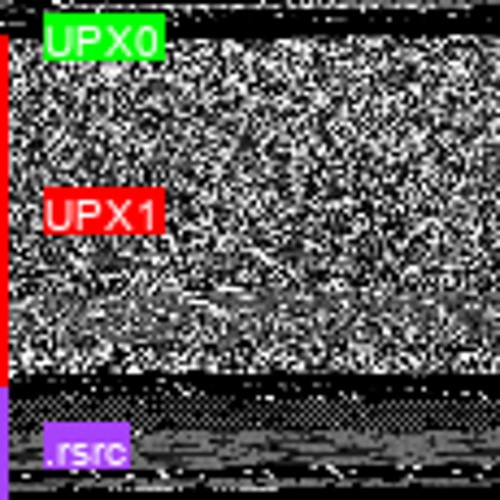}
	\label{Binary showing unusual features (zipped by UPX)}}
	\caption{Benign binary data (a); Unusual binary (b)}
	\label{example} 
\end{figure}

Examples of common and unusual binaries images are shown in Figure \ref{example}. Ransomware binaries images often show particular patterns that may be discovered by machine learning models:
\begin{itemize}
    \item Obfuscation methods may alter the binary sections in a recognizable way (like big sections with high entropy due to the payload being compressed),
    \item They behaviour are not so diverse (encrypt files, send them over the network for double-extortion, display instructions about the ransom payment). This is clearly reflected in their structure.
\end{itemize}

\subsection{Proposed CNN model for classification }


A Convolutional Neural Network Model is developed to classify the two classes. The CNN has a total of 3 hidden layers with 1 convolution layers, 1 Dropt Out layer and two fully connected layers. The final output layer is two classes, a classification layer with Sigmoid activation. The structure of CNN is shown in Figure \ref{cnn}, which contains a combination of convolutional layers, Dropout layer to prevent overfitting,  rectified linear unit (ReLU) activation layers, Fully connected layers and an output layer. The output feature map of each convolution in the three layers is applied to the ReLU activation function in the network \cite{agarap2018deep}. The ReLU function is defined
\begin{align}
  f({{z}^{k}})=\max \text{ }\{\text{ }0,{{z}^{k}}\text{ }\}= \begin{cases}
   & {{z}^{k}},{{z}^{k}}\ge 0 \\ 
    & 0,else, \\ 
\end{cases}
\end{align}

where ${z}^{k}$ is element of outputs in $k$th convolutional layer. 

The feature map is converted from a 2-dimensional matrix into a single vector. The Fully-connected layer is applied in the final convolutional layer with sigmoid activation function as shown:
\begin{align}
    Z={{\text{ }\!\![\!\!\text{ }{{z}_{1}},...,{{z}_{m}}\text{ }\!\!]\!\!\text{ }}^{T}}=\sigma (h)
\end{align}
where ${z}_{m}$ is the predicted fault type in the mth category in the M classes, $h={{\text{ }\!\![\!\!\text{ }{{h}_{1}},....,{{h}_{m}}\text{ }\!\!]\!\!\text{ }}^{T}}$
where $\sigma(h)$ is the sigmoid function, which is defined as:
\begin{align}
{{z}_{m}}={{\text{ }\!\![\!\!\text{ }\sigma (h)\text{ }\!\!]\!\!\text{ }}_{m}}=\frac{{{e}^{{{h}_{m}}}}}{\sum\nolimits_{j=1}^{M}{{{e}^{{{h}_{j}}}}}},
\end{align}

We conducted grid search experiments with different parameters to adjust our model to achieve the other optimized hyperparameters, such as the number of epochs, batch size, and learning rate. The CNN model structure of the proposed architecture in our work is detailed in Table \ref{table2} which contains name layers, output shapes, activation function, kernel number and padding. 

\begin{table*}
\caption{Overview of the Proposed Network Architecture}
\label{table2}
\setlength{\tabcolsep}{3pt}
\centering
\begin{tabular}{|c|l|p{1.7cm}|c|c|c|}
\hline
\textbf{No}& \textbf{Layer type}& \textbf{Kernel size/Stride}&\textbf{Kernel number}&\textbf{Output size}&\textbf{Padding}\\
\hline
1&Convolution + Relu &3x3/1&32&300x300x32&same\\
2&Drop out&-&-&300x300x32&-\\
3&Flatten layer  &-&4&2,880,000&same\\
4&Dense&-&2&-&\\

\hline
\end{tabular}
\label{tab1}
\end{table*}

\subsection{Federated Learning using CNN model}
The reason that a CNN model is used to implement federated learning is that CNN models can extract useful features and learn representations from images.
In the federated learning task, the CNN model is trained using the clients' local data. After the CNN model is trained locally, the weight of the model is aggregated to update the global model.

Assume that there are three people using computers: Bob, Alice and Sue. Our ultimate goal is to train a global model that can detect ransomware on Bob, Alice, and Sue computers. However,  the data that Bob, Alice, and Sue have is different. Bob may have a collection of ransomware images that Alice and Sue do not have. He may not want to share them. Besides, the data may contain sensitive and private information. For example, the file name or the path of the file may contain personal information. 

Because of the above reason, a federated learning task is needed to train the global model.

\subsection{Experiments}
\subsubsection{Results with CNN model}


In order to detect ransomware attacks, the CNN model needs to be trained with the prepared data. In this simulation experiment process, to achieve accurate results, image classification is the main result. The proposed CNN models are described as learning a target function that maps input variables to an output variable. This is a general learning task where predictions of the events will be executed using given examples of input variables. Therefore, the CNN will learn based on the input data.

There are two classes of data: normal and ransomware data.  The dataset in the experiments is shown in Table \ref{Experimental} together with the amount of data. The dataset was composed of a total of 6,000 samples that contain the two considered categories of "normal" and ransomeware. 

\begin{table}
	\caption{Experimental Dataset}\label{Experimental}
	\centering
	\begin{tabular}{|c|c|c|}
		\hline
		\textbf{Type of data} & \textbf{Normal (0)} & \textbf{Ransomeware (1)}\\
		\hline
		Number of sample & $3000$ & $3000$\\
		\hline
	\end{tabular}
\end{table}

Each class is assumed to be 0 and 1, corresponding to the normal and ransomware data. Then, we divided the data into three parts: 80 \% for training, 10 \% for validation and 10 \% for testing.  In the training phase, the optimization step was set to obtain the optimized hyper parameters corresponding to the layer type, batch size, and~a number of filter sizes. The~mini batch size is set as 64, the~learning rate is set as 0.006, and~the epoch is 10 for the training~model.

\begin{figure}[t]
    \centering
    \includegraphics[width=.7\textwidth]{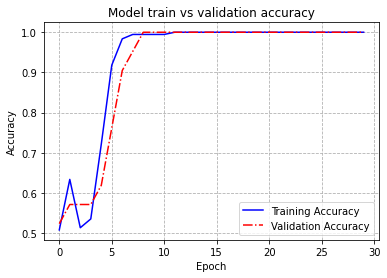}
    \caption{Traning and validation accuracy}
    \label{accuaracy}
\end{figure}

Figure~\ref{accuaracy} illustrates the training accuracy and validation accuracy over epoch number for the classification. In the initial stage, we observe that our model has a performance decrease on the training set. This is due to the initial training will only perform a simple approximation of the model. After about 4 epochs, the model's performance increases in both training set and the validation set. The~training accuracy and the validation accuracy were increased and converged to 100 \%.

\begin{figure}[t]
    \centering
    \includegraphics[width=.6\textwidth]{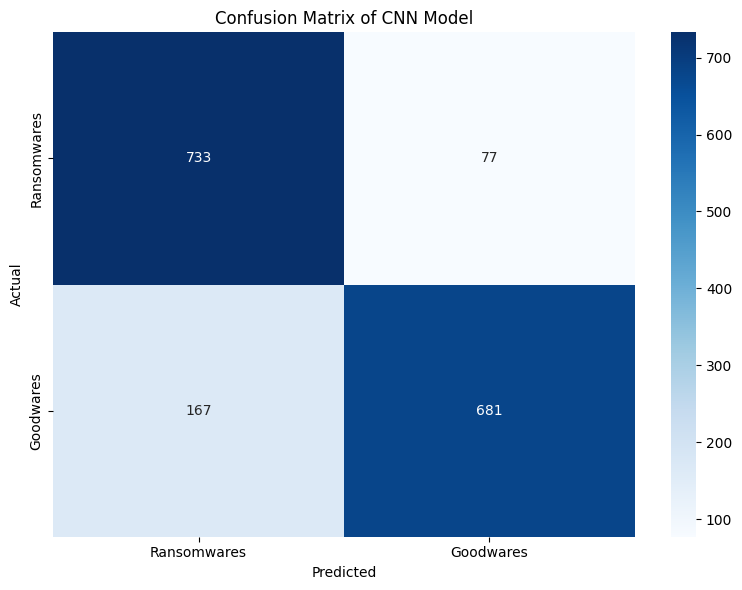}
    \caption{Confusion matrix of CNN model}
    \label{confusion}
\end{figure}

After training the prepared dataset with the CNN model, we evaluated the performance classification with the testing dataset. Figure \ref{confusion} presents the confusion matrix of CNN model when predicting on the testing dataset. 
These results validate that the proposed CNN  can successfully identify cyber-attacks with high accuracy based on the image feature.


\subsubsection{CNN model using Federated Learning} 

The purpose of federated learning is not to increase accuracy but to increase data privacy and solve the problem of data decentralization, so in this experiment, we simulate training data on different servers. Considering a real-world scenario where the training data for ransomware attacks is on  three different servers, this data is the combination of user behavior data and the data generated by the malware itself. To train a model that is more robust to ransomware detection, data from all three servers have to be used to train the model. However, data from each server is sensitive and should not be shared with other servers. So the federated learning concept can be used to train a model that is more robust to ransomware detection. In federated learning, the model is trained on each server with its own data and then the weights of the model are shared with the other servers. In this way, the data remains private.


We performed comprehensive federated learning experiments with different parameters to adjust the CNN model to achieve the other optimized hyperparameters, such as the number of clients, the number of epochs, batch size, and the learning rate. At the end of the turning process, we have a combination of the parameter as shown in Table \ref{prFedrate}



Table \ref{classificationreport3} shows the precision, recall, and F1-score for the CNN model using federated learning. 
For precision and recall, the detection scores for normal and Ransomeware attacks achieved identical scores of 92\% and 100\%, respectively. Similarly, recall scores of 100\% and 92\%, respectively, are obtained for the normal and Ransomeware attacks. The F1-Score is 96\%, which proves the effectiveness of the proposed CNN model using Federated learning.

\begin{table}
	\caption{Classification Report of Federated Learning }\label{classificationreport3}
	\centering
		\begin{tabular}{|c|c|c|c|c|}
		\hline 
		\textbf{Classes}	 & \textbf{Precision}& \textbf{Recall}& \textbf{F1 score}\\\hline
		\textbf{Goodware} & $81 \%$ & $90\%$ & $86 \%$\\
    	\hline
	 	\textbf{Ransomeward} & $90 \%$ & $80 \%$ & $85 \%$ \\
	    \hline	
	\end{tabular}
\end{table}

\begin{table}           
    \caption{Hyperparameter Federated learning} \label{prFedrate}
    \centering
    \begin{tabular}{|c|c|}
    \hline
    \textbf{Hyperparameter} & \textbf{Value}\\
    \hline
    Method  & "tff\_training"\\
    \hline
    Number of clients &$3$\\
    \hline 
    NUM\_ROUNDS & $30$ \\
    \hline
    NUM\_EPOCHS &$30$\\
    \hline
    BATCH\_SIZE & 64\\
    \hline
    \end{tabular}
\end{table}


\subsubsection{Discussions}
Although the experimental results show some positive results, there are some limitations to our experimental setup. First, the size of our dataset is relatively small. This limitation may lead to overemphasizing the robustness of our federated learning algorithm. Future work will consider a larger dataset. Second, our dataset is mostly equally distributed among clients. However, the data is distributed unequally among clients in the real world. We will consider this case in future work. Third, our experiment can easily split the data into two separate labels, normal and abnormal. However, real-world data may be far more complex and have more classes. We will also consider this case in future work. 

Also, there is a high risk to learn two important bias. Ransomware binaries are often quite small and zipped by UPX to escape signature based detection. That's why we have to be particularly careful when creating the learning dataset. 

\section{Conclusions}

The Proposed system can classify normal and ransomware attacks with high efficiency. The Proposed CNN Model has been tested with a variety of federated learning techniques and the attention of the CNN has been visualized for a better understanding of the classification decisions made by the model to differentiate. Moreover, there has been a detailed analysis done for different Fedarate learning training techniques for the model and further, the model layers and training parameters can be tweaked to improve and make it more robust.

\renewcommand{\bibsection}{\section*{References}}

\bibliographystyle{splncs04}
\bibliography{ref}
\end{document}